# Power-like Tail Observed in Weight Distributions of Schoolchildren

Hiroto Kuninaka

*Faculty of Education, Mie University, Tsu City, Mie 514-8507, Japan*

We investigated the statistical properties of the weight distributions of Japanese children who were born in 1996, from recent data. The weights of 16- and 17-year-old male children have a lognormal distribution with a power-like tail, which is best modeled by the double Pareto distribution. The emergence of the power-like tail may be attributed to the low probability that an obese person will attain a normal weight.

Height and weight are important indices for human development. Weight in particular is often an object of concern because it is closely related to health condition. Recently, the McKinsey Global Institute reported that nearly 30 percent of the world population suffers from obesity and overweight, which can increase the incidence of cardiovascular diseases, heart failure, and diabetes[1]. Obesity is nowadays considered to be a kind of epidemic[2], and the measurement of abdominal circumference and the calculation of Body Mass Index (BMI) have been required items in health checks in Japan since 2008. This reflects the fact that obesity and metabolic syndrome are becoming serious problems among Japanese adults.

Height and weight distributions are useful for capturing the health condition of residents in a given region. Although it is generally believed that those distributions are well approximated by the normal distribution, we often find skewed distributions in the case of height and weight of growing children under age 20. For example, Mitsuhashi et al. analyzed height and weight data from 1953 to 2006 to investigate the time evolution of the height and weight distributions of Japanese schoolchildren[3]. They found that the height distributions could be better approximated by the lognormal distribution than by the normal distribution before the onset of puberty[3-6]. In addition, they found that the cumulative distributions of the weight $x$ of Japanese schoolchildren aged 5 to 17 could be well approximated by a mixture of cumulative lognormal distributions defined by

$$F(x) = \sum_{i=1}^{N} w_i f_i(x), \quad \text{where} \quad \sum_{i=1}^{N} w_i = 1, \tag{1}$$

while those of newborn babies could be approximated by a single lognormal distribution[3,7]. Here, $f_i(x)$ is a cumulative lognormal distribution,

$$f_i(x) = \frac{1}{2}\left[1 - erf\left\{\frac{\log(x/\mu_i)}{\sqrt{2}\sigma_i}\right\}\right] \qquad (2)$$

with the error function $erf(x)$ weighted by $w_i$. Mitsuhashi et al. also found that the number of components changed from $N = 3$ to $N = 2$ with increasing age[3].

In this short note, we report some recent analyses of weight data of Japanese schoolchildren. We analyzed data of the school health survey by the Ministry of Education, Culture, Sports, Science and Technology (MEXT) from 2007 to 2013[8]. This survey is carried out yearly for the schoolchildren aged 5 to 17 who belong to schools randomly chosen from all the schools in Japan. In 2014, for example, the numbers of sampled schools and children were 7,755 and 695,600, respectively. Note that the data obtained from MEXT are binned histogram data. In the case of 12-year-old male students in 2014, for example, all the sampled weights are split into 94 equally sized (1 kg) classes ranging from 20 to 117 kg. Thus, the interpretation of the data is subject to sampling errors because of the sample survey. We collected weight data of Japanese male children born in 1996 from 2007 (age 11) to 2013 (age 17). Instead of the density distribution, we investigated the cumulative weight distribution each year to smooth the noisy profile of the density distributions.

Figure 1 shows the log-log plot of the cumulative weight distribution of 16-year-old male children in 2012, where the horizontal and vertical axes respectively show the logarithm of the weight scaled by 61 kg and the logarithm of the frequency scaled by total sampling numbers. The dotted curve shows the double lognormal cumulative distribution (dLN), the case of N=2 in Eq. (1), which approximates the data. For fitting the data, we used gnufit, which is a nonlinear least squares fit program implemented in gnuplot. Note that the tail shows a power-like shape, so that the discrepancy between the data and the approximating distribution becomes large in the tail. We found a similar tendency in the case of 17-year-old male children in 2013. The cumulative weight distributions of male children before age 15 are well approximated by Eq. (1), and the number of components changes from N=3 to N=2, which is consistent with the analysis by Mitsuhashi[3]. The empirical distribution with the lognormal body and the power-law tail reminds us of the double Pareto distribution (dP) advocated by Reed et al.[9], which has been observed in the file size distribution in hard disks[10] and the prefectural population distribution in Japan[11]. The cumulative form of dP is described as

$$R(x) = \begin{cases} 1 - \dfrac{a}{a+b}(x/c)^b & (0 \leq x/c < 1) \\ \dfrac{b}{a+b}(x/c)^{-a} & (x/c \geq 1) \end{cases} \quad (3)$$

with parameters $a$, $b$, and $c$ [10]. The solid curve in Fig. (1) shows Eq. (3) with $a = 7.92515$, $b = 6.75382$, and $c = 61$ kg. In this case, as in the case of the dLN model, the discrepancy between the data and Eq. (3) becomes large in the tail. In order to judge which distribution is a better model for the weight distributions of 16-year-old and 17-year-old male children, we calculated the Akaike information criterion (AIC)[12] for each fitting function according to the definition,

$$AIC = -2\ln M + 2k, \quad (4)$$

where $M$ and $k$ are the maximum likelihood and the number of parameters, respectively. In Table I, calculated AICs are summarized, where tLN shows the triple lognormal distribution, i.e. Eq. (1) with N=3. This result shows that the best model for the data is the double Pareto distribution, in spite of the large discrepancy in the tail.

Let us consider the origin of the power-like tail in the weight distribution. The tail includes tall children in addition to obese children. In the case of male children taller than 187 cm, about 3 ‰ of all children at age 16, those weighing more than 85.08 kg are categorized as obese[8, 13]. Because the tail starts around at 100 kg, we can conclude that most children in the tail are obese.

We may be able to assume the development of weight as a random multiplicative process because the weight distributions are composed of lognormal distributions. On the other hand, it is known that a random multiplicative process can generate a power-law distribution under the condition that the size of a developing body does not fall under a lower bound[14]. Very recently, Fildes et al. reported that the probability that an obese person, aged from 20 years and sampled from UK's Clinical Practice Research Datalink from 2004 to 2014, will recover normal weight is extremely low, so that it is very hard for obese people to attain normal weight once they are categorized as obese[15]. Thus, also in the case of Japanese children who are close to age 20, we conclude that the criterion of obesity acts as a kind of lower bound so that the tail part of the distribution shows a power-like shape. A similar emergence mechanism of a power-like tail can be found in the formation of the population distribution of municipalities in Japan[11].

In conclusion, we report the statistical characteristics observed in the weight distribution of schoolchildren from recent data. The weight distribution of ages 16 and 17 have a lognormal distribution with a power-like tail, whereas the weight distributions of ages under 15 change from tLN to dLN with increasing age, confirming previous results by Mitsuhashi et al. Although we believe that the origin of the power-like tail is the low probability that an obese child will recover a normal weight based on a recent research, we found lognormal distributions with a power-like tail in the weight data of female children aged 6, 8, and 11 in 1953[3]. Thus, in our future work, we will analyze recent weight data of female children in more detail, as well as the body size data of other countries such as the USA , where obesity is a more serious health problem.

**Acknowledgement(s)**

This work was supported in part by a Grant for Basic Science Research Projects from the Sumitomo Foundation.

*E-mail: kuninaka@edu.mie-u.ac.jp

**figure caption(s)**

Fig. 1: Log-log plot of cumulative weight distribution of 16-year-old male children in 2012. Solid and dotted curves are double Pareto and double lognormal distributions, respectively.

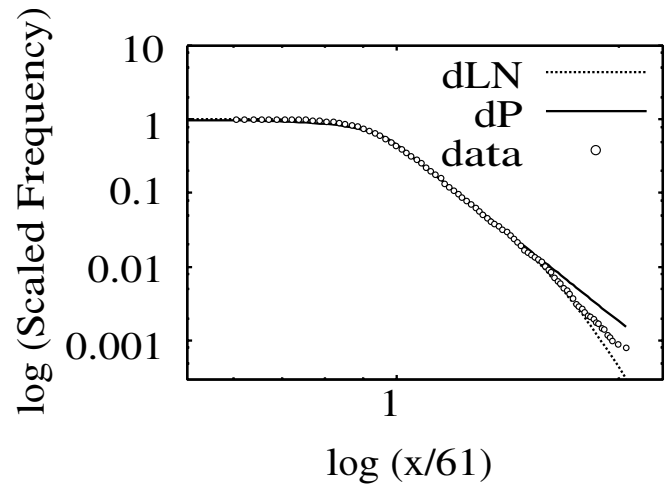

Fig. 1

Table I: AICs calculated with three kinds of statistical models.

| Model | Age 16 | Age 17 |
|---|---|---|
| dLN (N=2) | 538.2604 | 523.0849 |
| tLN (N=3) | 534.7753 | 519.1113 |
| dP | 489.0202 | 480.9081 |